\documentstyle[prb,aps,psfig]{revtex}

\begin{document}
\title{A search for incipient lattice instabilities in MgB$_{2}$ by anelastic
spectroscopy}
\author{F.Cordero,$^{1}$ R. Cantelli,$^{2}$ G. Giunchi,$^{3}$ and S. Ceresara$^3$}
\address{$^{1}$CNR, Istituto di Acustica ``O.M. Corbino``, Via del Fosso del
Cavaliere 100,\\
I-00133 Roma, Italy and Unit\`{a} INFM-Roma1, P.le A. Moro 2,I-00185 Roma,
Italy}
\address{$^{2}$Universit\`{a} di Roma ``La Sapienza``, Dipartimento di Fisica, and\\
Unit\`{a} INFM-Roma 1, P.le A. Moro 2, I-00185 Roma, Italy}
\address{$^{3}$ EDISON S.p.A, Foro Buonaparte 31, I-20121 Milano, Italy}
\maketitle

\begin{abstract}
We measured the complex dynamic elastic modulus of MgB$_{2}$ between 1.3 and
650~K at frequencies included between 5 and 70~kHz. The main purpose of this
work is to search for possible lattice instabilities, which could contribute
to enhance the electron-phonon coupling. The Young's modulus indeed presents
an anomalous softening on cooling below 400~K, and hysteresis between
cooling and heating. Such anomalies, however, can be accounted for by
intense relaxation processes with maxima between 50 and 150~K, whose nature
is not yet clarified, and by geometric defects generated by the anisotropic
thermal expansion between the grains during thermal cycling. Additional
intense relaxation processes are observed above room temperature, and their
possible origin is discussed. The influence of the highly anharmonic
in-plane vibration modes of the B atoms on the elastic and anelastic
properties of MgB$_{2}$ is discussed in detail.
\end{abstract}

\twocolumn

The discovery of superconductivity with $T_{\text{c}}=$ 39~K in MgB$_{2}$ is
arousing great interest, since this intermetallic compound seems to be a
simple BCS superconductor, without any of the other mechanisms for high-$T_{%
\text{c}}$ superconductivity that are hypothesized for the cuprate
superconductors. Electronic band calculations\cite{AP01,YGL01} indicate a
predominant hole conduction in the two-dimensional $\sigma $ band from the $%
sp_{x}p_{y}$ orbitals of B, with an extremely large deformation potential
for the B bond stretching mode, which dominates the electron-phonon
coupling. The most recent first principle calculations of the electronic
bands and phonons\cite{YGL01} indicate that the zone-center optical mode
with E$_{2g}$ symmetry, involving vibrations of the B atoms within the $ab$
plane, has very flat anharmonic potential, and can account for the observed $%
T_{\text{c}}$ with a calculated electron-phonon coupling constant $\lambda
\simeq 1$. An experimental indication of the anharmonicity of that mode
comes from the high value of its Gr\"{u}neisen parameter, evaluated from the
shift of the corresponding Raman band with pressure.\cite{GSG01} Still, it
is important to ascertain whether there are other anharmonic or unstable
modes that are not easily seen by phonon DOS measurements\cite{YGL01} and
other spectroscopies, but can contribute to the large electron-phonon
coupling. In many cases of BCS superconductivity with high $T_{\text{c}}$,
the strongly anharmonic phonons causing the enhanced electron-phonon
coupling also produce a structural instability, which in general profoundly
affects the acoustic properties of the material. The best known examples are
the A15 superconductors,\cite{Tes73} where a clear correlation exists
between the occurrence of a transformation from cubic to tetragonal, often
assimilated to a martensitic transformation, and a high value of $T_{\text{c}%
}$. It should be emphasized that the occurrence and the entity of the
structural instability in the A15 compounds is rather sensitive to the
material preparation, and the easiest method for revealing it was the
observation of softening of the elastic moduli on cooling below room
temperature.\cite{Tes73} Also the superconducting perovskite BKBO presents
structural instabilities,\cite{ZLG00,BRE00} while ultrasonic absorption
peaks in BPBO are interpreted in terms of strong phonon anharmonicity.\cite
{MFH}

In the case of MgB$_{2}$ a possible unstable mode is the B$_{1g}$ mode of
the out-of plane vibrations of the B atoms, even though the first principle
calculations indicate a stiff and harmonic mode.\cite{YGL01} In fact, it has
been observed that MgB$_{2}$ is close to the condition of buckled B planes
in a phase diagram of the AB$_{2}$ compounds versus charge density and B-B
distance.\cite{BSC01} In addition, when the hole density in the B planes is
progressively reduced by partially substituting Mg with Al, the $c$ lattice
parameter collapses in correspondence with the disappearance of
superconductivity, suggesting that MgB$_{2}$ is close to a structural
instability at slightly higher electron density.\cite{SRR01}

The anelastic spectroscopy is very sensitive to structural instabilities,
and therefore we measured the complex dynamic Young's modulus of MgB$_{2}$
between 1.3 and 650~K, searching for possible signs of lattice instabilities
not yet detected by neutron diffraction\cite{YGL01,JHS01} and other
spectroscopies.

The sample was prepared by reacting Mg and B in a sealed stainless steel
tube, lined with Nb foil, at 950~$^o$C for 2 hours. Further preparative
details are reported elsewhere.\cite{Edison} A bar $39\times 6\times 1$~mm$%
^{3}$ was cut by spark erosion. The mass density was $\rho =2.27$~g/cm$^{3}$%
, 86\% of the theoretical one. The superconducting transition, measured by
resistivity, was centered at 39.5~K, with a width of 1~K. The X-ray
diffraction spectra showed the presence of a residual minority phase of
metallic Mg, presumably at the grain boundaries. The sample was suspended
with thin thermocouple wires and electrostatically excited on its flexural
modes, whose resonant frequencies are given by $f_{n}=\alpha _{n} h/l^{2} 
\sqrt{E^{\prime }/\rho }$, where $E\left( \omega ,T\right) =E^{\prime
}+iE^{\prime \prime }$ is the complex dynamic Young's modulus,\cite{NB} $l$
and $h$ the sample length and thickness, and $\alpha _{n}$ numerical
constants depending on the vibration mode. The 1st, 3rd and 5th modes could
be excited, at $\omega /2\pi =$ 5.7, 38 and 73~kHz respectively at 1~K. The
value of the Young's modulus at room temperature, not corrected for the
sample porosity and the Mg traces, is $E=167$~GPa. The imaginary part is
related to the elastic energy loss coefficient $Q^{-1}=$ $E^{\prime \prime
}/E^{\prime }$, which was measured from the width of the resonance curves.

Figure 1 shows the anelastic spectrum between 1 and 620~K, measured by
exciting the first flexural mode. A first result is the absence of anomalous
softening below 150~K, so excluding a scenario like that of the A15
compounds, where the strong phonon anharmonicity is due to a mode which
becomes unstable slightly above $T_{\text{c}}$. Still, anomalous softening
and hysteresis between cooling and heating are observed between 400 and
150~K. Such anomalies, however, should not be straightforwardly attributed
to a structural instability, since microplastic phenomena connected with the
anisotropic thermal expansion of the hexagonal crystallites are expected.
Neutron diffraction\cite{JHS01} experiments have shown a large anisotropy of
the thermal expansion and compressibility, both being about twice larger
along the $c$ axis. These plastic phenomena have already been observed in
other hexagonal polycrystals, like Zn, Co and Cd,\cite{YB93} whose thermal
expansivity is sufficiently anisotropic and the critical stress $\sigma _{c}$
for plastic deformation sufficiently low to yield the formation of
dislocations during thermal cycling. The increase of the density of
geometrical defects results in a reduction of the modulus, and the
competition between the defect generation and annihilation, the latter
especially at high temperature, explains the hysteresis between cooling and
heating above 130~K. Below that temperature the thermal expansion becomes so
small that the stress generated at the grain boundaries is below $\sigma
_{c} $ and no further modification in the geometrical defects occurs. We do
not know what type of geometrical defects accommodates the strains generated
during thermal cycling in MgB$_{2}$, but they are not exclusively
accommodated at the Mg impurities between the grains; in fact, after heating
the sample up to 690~K in vacuum, about 8\% of Mg was lost, the absorption
component at 80~K was reduced, but the hysteresis of the modulus below room
temperature was unchanged. Also the elastic energy loss is partially due to
the motion of the geometrical defects, at least above 150~K, since it\ has a
component which increases with the cooling/heating rate, and exhibits
hysteresis, as shown by the open symbols in Fig. 1. The plastic phenomena
can account for the irregularities and hysteresis of $E\left( T\right) $
between 150 and 400~K but probably not completely for the plateau of $%
E^{\prime }\left( T\right) $ between these temperatures. This plateau could
also have another origin; indeed, a small softening of some modes appears in
the phonon DOS measured by neutron scattering\cite{YGL01} around 325~K. We
shall see however, that also the relaxation processes observed below room
temperature in the elastic energy loss might cause a similar effect.

The intense absorption peak at 475~K is due to a thermally activated
relaxation process,\cite{NB} namely it can be described as a contribution $%
\delta E=-\Delta /\left( 1+i\omega \tau \right) $ to $E\left( \omega
,T\right) $, where the relaxation time follows the Arrhenius law $\tau =\tau
_{0}\exp \left( W/T\right) $. The absorption peak and modulus dispersion of
this type of processes shift to higher temperature with increasing
frequency. Due to the high damping, the higher frequency modes could be
followed only up to 440~K, but exhibited such a shift, and could be
interpolated by the expression 
\begin{equation}
\delta E=-\Delta /\left[ 1+\left( i\omega \tau \right) ^{\alpha }\right] \,,
\label{FK}
\end{equation}
where the parameter $\alpha =0.65$ reproduces the broadening of the spectrum
of the energy barriers ($\alpha =1$ in the monodispersive case), $W=0.89$~eV
and $\tau _{0}=10^{-14}$~s. The dotted lines in Fig. 1 are the corresponding
contributions to the absorption and to $E^{\prime }$, plus the tail of
another process at higher temperature, probably connected with
grain-boundary motion or with the loss of Mg. The value of $\tau _{0}$ is
typical of point defect motion, as could be the migration of Mg atoms and
vacancies, but, due to the bad quality of the higher frequency data, we
cannot exclude that $\tau _{0}$ is actually longer and therefore connected
with the motion of geometrical defects.

In view of the occurrence of plastic phenomena during thermal cycling above $%
\sim 130$~K, it is likely that the absorption contains an important
contribution from the relaxation of extended lattice defects. Some
contribution may also be expected from Mg at the grain boundaries, since
deformed Mg presents an extremely broad absorption maximum\cite{TS67}
reminiscent of the curves found here below room temperature; nevertheless,
the contribution of metallic Mg is weighted with its molar fraction, i.e.
few percents, and considering that the intensity of the peak in deformed Mg%
\cite{TS67} never exceeds $2\times 10^{-3}$, it cannot account for elastic
energy loss curves of Fig. 2. In addition, at $T_{\text{c}}$ it is evident a
change of the slope of both the imaginary and real parts of $E\left(
T\right) $, which must be connected with processes occurring in the
superconducting bulk of MgB$_{2}$. Another unusual feature of the anelastic
spectrum below 200~K is that the curve simultaneously measured at higher
frequency, instead of being simply shifted to higher temperature, as for any
thermally activated relaxation process, is about a factor 1.32 more intense
than that at lower frequency. Finally, it should be noted that the spectrum
below 100~K is perfectly reproducible during various cooling and heating
cycles; considering the extreme sensitivity of the dislocation peaks to the
sample state, and the occurrence of plastic phenomena during thermal
cycling, it is unlikely that the low temperature spectrum is exclusively
connected with geometrical defects.

It is tempting to relate the low temperature anelastic spectrum with the
highly anharmonic E$_{2g}$ optical mode which is thought responsible for the
strong electron-phonon coupling. Indeed, a relaxation mechanism involving
the soft optical modes responsible for the structural phase transformations
in perovskites has been proposed by Barrett,\cite{Bar70} and later adopted
in order to explain the ultrasonic absorption in the perovskite
superconductor\cite{MFH} BaPb$_{1-x}$Bi$_{x}$O$_{3}$ and even in some
cuprate superconductors.\cite{HTM89} The mechanism is analogous to the
Akhiezer-type absorption,\cite{NB} but instead of the acoustic phonons, it
is based on the existence of optical phonons with a large Gr\"{u}neisen
constant, $\gamma =-d\left( \ln \omega _{0}\right) /d\varepsilon $, namely
whose frequency $\omega _{0}$ strongly depends on a strain $\varepsilon $.
The flexural vibration of the sample causes a non homogeneous uniaxial
strain $\varepsilon $ along the sample length and modulates the energies of
the phonons according to their Gr\"{u}neisen parameters. The phonons system
goes out of thermal equilibrium if different modes have different values of $%
\gamma $, and the equilibrium is established with an effective phonon
relaxation time $\tau _{\text{eff}}$. The change of the population is
reflected in the dynamic elastic modulus, which acquires an out of phase
imaginary component and a temperature dependent softening. It is appropriate
to discuss this mechanism, since the Gr\"{u}neisen constant of the E$_{2g}$
optical mode has been found to have a very large value at room temperature,%
\cite{GSG01} $\gamma _{\text{op}}=\frac{1}{3}\frac{d\ln \nu }{d\ln a}\simeq
3.9$, $a$ being the lattice constant, while normal values range between 1
and 2. Barrett solved the case in which the system is schematized with an
optical mode with frequency $\omega _{0}$ and large $\gamma _{\text{op}}$,
and the acoustic modes with normal $\gamma _{\text{ac}}$; since we are
considering uniaxial strain $\varepsilon =da/a$ instead of volume change, we
define $\overline{\gamma }=3\left( \gamma _{\text{op}}-\gamma _{\text{ac}%
}\right) \simeq 7$, where $\gamma _{\text{ac}}\sim $ 1.5 is assumed. The
contribution of the phonon relaxation to the dynamic Young's modulus $E$ is%
\cite{Bar70} 
\begin{eqnarray}
\delta E &=&-\frac{\overline{\gamma }^{2}}{v_{0}}TC_{\text{op}}\frac{%
1+i\omega \tau _{\text{eff}}}{1+\left( \omega \tau _{\text{eff}}\right) ^{2}}
\\
\tau _{\text{eff}} &=&\tau _{\text{op}}+\frac{C_{\text{op}}}{C_{\text{ac}}}%
\tau _{\text{ac}}
\end{eqnarray}
where $v_{0}$ is the cell volume, $C_{\text{ac}}$ and $C_{\text{op}}$ are
the specific heat of the acoustic and optic modes per molecule and $\tau _{%
\text{ac}}$ and $\tau _{\text{op}}$ are their relaxation times. The elastic
energy loss coefficient is, for $\omega \tau _{\text{eff}}\ll 1$, 
\begin{equation}
Q^{-1}=\frac{E^{\prime \prime }}{E^{\prime }}=\frac{\overline{\gamma }^{2}}{%
v_{0}E}TC_{\text{op}}\omega \tau _{\text{eff}}=\Delta \left( T\right)
\,\omega \tau _{\text{eff}}\,;
\end{equation}
where $C_{\text{op}}$ has the usual expression $k_{\text{B}}\left[ \left(
\hbar \omega _{0}/2T\right) /\sinh \left( \hbar \omega _{0}/2T\right)
\right] ^{2}$ with\cite{GSG01} $\hbar \omega _{0}/k_{\text{B}}=830$~K, $%
E=167 $~GPa and $v_{0}=29\times 10^{-24}$~cm$^{3}$; the resulting relaxation
strength $\Delta $ is $6.6\times 10^{-5}$ at 100~K and $6.5\times 10^{-3}$
at 300~K, but has to be multiplied by the factor $\omega \tau _{\text{eff}}$%
, which is very small. In fact, the phonon relaxation rates and therefore $%
\tau _{\text{eff}}^{-1}$ should exceed $10^{12}$~s$^{-1}$; $\tau _{\text{ac}%
} $ can be deduced from the lattice thermal conductivity $\kappa $ through
the collision formula $\kappa =\frac{1}{3}C_{\text{ac}}\overline{v}^{2}\tau
_{\text{ac}}$. Assuming $\kappa \simeq $ 0.1~W/(cm~K) found\cite{BPB01}
above 100~K, an average sound velocity somewhat lower than that deduced from 
$\sqrt{E/\rho }=$ $8\times 10^{5}$~cm/s, and a Debye lattice specific heat,
it results $\tau _{\text{ac}}<10^{-12}$~s; $\tau _{\text{op}}$ can be
assumed to be of the same order of magnitude or even smaller. Since our
measuring frequency is only $\omega \sim 10^{4}-10^{5}$, the $\omega \tau _{%
\text{eff}}$ factor makes the imaginary part negligibly small, and therefore
the absorption peaks around 100~K cannot be attributed to an Akhiezer type
mechanism involving the anharmonic optic phonon. Still, the real part of $%
\delta E$ contributes to the modulus softening with a substantial $\delta
E/E=-\Delta \left( T\right) $; the continuous curve in Fig. 1 is calculated
with the above parameters, assuming a temperature independent $\overline{%
\gamma }=7$.

Although the mechanisms producing the acoustic absorption below 150~K remain
unclear, it is worth checking whether they can also be the origin of the
anomalous softening observed below 400~K. The imaginary part of the dynamic
modulus should be describable in terms of a superposition of relaxation
processes like Eq. (\ref{FK}), where the phenomenological parameter $\alpha
\le 1$ describes the broadening of the spectrum of relaxation times around
an average value $\tau $, and the relaxation strength is generally $\Delta
\propto 1/T\;$or nearly constant. In order to reproduce the experimental
curves, we need a $\Delta \left( T\right) $ strongly increasing with
temperature, and therefore we adopt $\Delta \left( T\right) \propto 1/\left[
T\cosh ^{2}\left( E/2T\right) \right] $, appropriate\cite{35} for relaxation
occurring between states which differ in energy by $E$. The solid lines
roughly interpolating the absorption data in Fig. 2 are obtained with the
following parameters, $\tau =$ $5\times 10^{-14}\exp \left( 800/T\right) $%
~s, $\alpha =0.2$, $E/k_{\text{B}}=230$~K for the peak at 70-80~K, and $%
5\times 10^{-15}\exp \left( 2900/T\right) $~s, $\alpha =0.4$, $E/k_{\text{B}%
}=130$~K for the peak at 150~K. We do not attribute a particular meaning to
these values, since it is possible that the two peaks are superimposed to a
broader maximum, which has not been considered. The fit is therefore only
indicative, but it shows that the anomalous softening below 400~K may be
accounted for by the same relaxation processes which cause absorption at low
temperature; in fact, the continuous line in the upper panel of Fig. 2 is
the real part corresponding to the curves in the lower panel and even
exceeds the experimental softening. The fact that the magnitude of the
negative step of $E^{\prime}$ is much larger than the amplitude of the $%
E^{\prime\prime}$ peaks is a consequence of the non additivity of the
amplitudes of the elementary peaks, whose maxima occur at different
temperatures.

The presence of these intense relaxation processes also masks the negative
jump and positive change of the slope of the elastic moduli due to the
superconducting transition;\cite{Tes73} in fact, the latter are of the order
of $10^{-4}$ or less, while the low temperature relaxation processes cause a
modulus defect of the order of $10^{-3}$ at 40~K, and they are affected by
the transition to the superconducting state, as observed in both the real
and imaginary parts of $E $. In order to analyze the intrinsic changes of
the moduli at $T_{\text{c}} $ one should be able to subtract the
relaxational contribution. For this reason, although we can see the expected
jump and change of slope of $E^{\prime }\left( T\right) $ below $T_{\text{c}}
$, we omit the discussion of this issue.

In conclusion, we measured the complex dynamic elastic modulus of MgB$_{2}$
with the principal aim of searching for possible lattice instabilities,
which might contribute to enhancing the electron-phonon coupling without
being yet revealed by other spectroscopies. The Young's modulus indeed
presents an anomalous plateau on cooling between 400 and 150~K, and
hysteresis between cooling and heating. The softening, however, can be
accounted for by intense relaxation processes with maxima between 50 and
150~K, whose nature is not yet clarified. The hysteresis is explainable in
terms of generation and annihilation of geometric defects due to the
stresses generated by the anisotropic thermal expansion between the grains
during thermal cycling. The influence of the highly anharmonic vibration
modes of the B atoms on the elastic and anelastic properties of MgB$_{2}$ is
discussed in detail; such anharmonic phonons produce negligible acoustic
absorption at the low frequencies of the present study, but they may
substantially contribute to the total softening of the elastic modulus with
increasing temperature.

Edison acknowledges the Lecco Laboratory of the CNR-TeMPE, Milano, Italy,
for making available its facilities for materials preparation.


\begin{figure}[]
\caption{Real and imaginary parts of the anelastic spectrum of MgB$_{2}$,
measured exciting flexural vibrations at 5~kHz. The arrows and numbers
indicate the order of the cooling and heating runs. The dotted lines in the
lower and upper panel are a fit to the high temperature relaxation peaks and
the corresponding modulus defect, respectively. The dashed line is the
estimated softening from the anharmonic E$_{2g}$ optic mode of the in-plane
B vibrations, with the value of the Gr\"{u}neisen parameter measured at room
temperature. The gray curve is the experimental Young's modulus after
subtraction of the above contributions.}
\label{fig1}
\end{figure}

\begin{figure}[]
\caption{Real and imaginary parts of the anelastic spectrum of MgB$_{2}$,
measured exciting flexural vibrations at 5.7~kHz and 73~kHz during the same
cooling run. The continuous lines in the lower panel are a fit assuming
broad thermally activated relaxation processes with the parameters given in
the text; the line in the upper panel is the corresponding contribution to
the real part of the elastic modulus. The arrows indicate the
superconducting $T_{\text{c}}$, at which the curves change the slope.}
\label{fig2}
\end{figure}


\begin{references}
\bibitem{AP01}  J.M. An and W.E. Pickett, Phys. Rev. Lett. {\bf 86}, 4366
(2001).

\bibitem{YGL01}  T. Yildirim {\it et al.}, cond-mat/0103469.

\bibitem{GSG01}  A.F. Goncharov {\it et al.}, cond-mat/0104042.

\bibitem{Tes73}  L.R. Testardi, {\it Physical Acoustics}. ed. by W.P. Mason
and R.N. Thurston, p. 193 (Academic, New York, 1973).

\bibitem{ZLG00}  S. Zherlitsyn {\it et al.}, Eur. Phys. J. B {\bf 16}, 59
(2000).

\bibitem{BRE00}  M. Braden {\it et al.}, Phys. Rev. B {\bf 62}, 6708 (2000).

\bibitem{MFH}  S. Mase {\it et al.}, Solid State Commun. {\bf 57}, 227
(1987).

\bibitem{BSC01}  A. Bianconi {\it et al.}, cond-mat/0102410.

\bibitem{SRR01}  J.S. Slusky {\it et al.}, cond-mat/0102262.

\bibitem{JHS01}  J.D. Jorgensen {\it et al.}, cond-mat/0103069.

\bibitem{Edison}  Edison S.p.A patent pending.

\bibitem{NB}  A.S. Nowick and B.S. Berry, {\it Anelastic Relaxation in
Crystalline Solids}. (Academic Press, New York, 1972).

\bibitem{YB93}  M.H. Youssef and P.G. Bordoni, Phil. Mag. A {\bf 67}, 883
(1993).

\bibitem{TS67}  R.T.C. Tsui and H.S. Sack, Acta metall. {\bf 15}, 1715
(1967).

\bibitem{Bar70}  H.H. Barrett, {\it Physical Acoustics}. ed. by W.P. Mason
and T.N. Thurston, p. 65 (Academic Press, New York, 1970).

\bibitem{HTM89}  Y. Horie {\it et al.}, J. Phys. Soc. Jap. {\bf 58}, 279
(1989).

\bibitem{BPB01}  E. Bauer {\it et al.}, cond-mat/0104203.

\bibitem{35}  F. Cordero, Phys. Rev. B {\bf 47}, 7674 (1993).
\end{references}
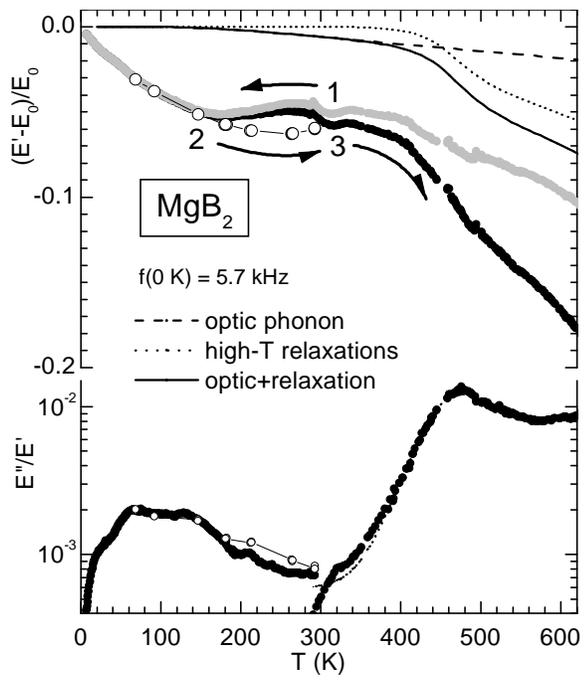
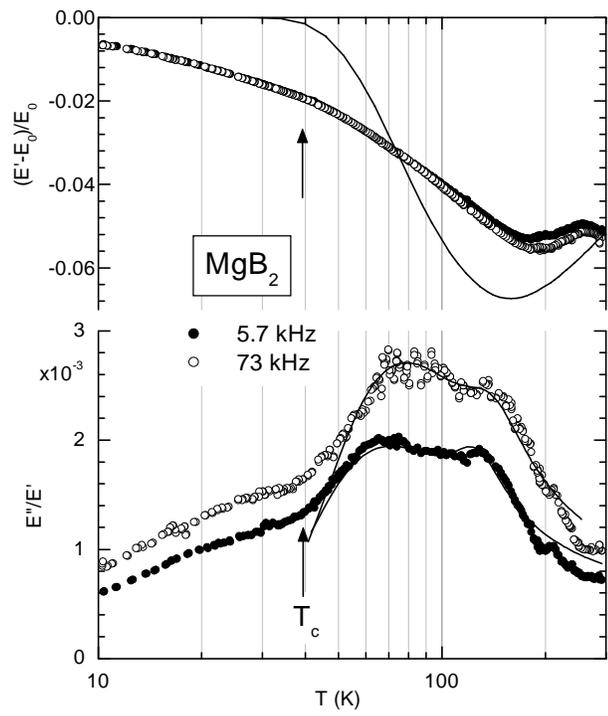
\end{document}